\documentclass[conference]{IEEEtran}
\usepackage{cite}
\usepackage{amsmath,amssymb,amsfonts}
\usepackage{algorithmic}
\usepackage{graphicx}
\usepackage{textcomp}
\usepackage{xcolor}
\usepackage{multirow}
\usepackage{booktabs}
\usepackage{diagbox} 
\usepackage[bookmarks=false]{hyperref}
\usepackage{tikz,forloop,tabularx,tabulary,multicol,multirow,pgfmath}
\usetikzlibrary{positioning,arrows,chains,matrix,scopes,fit,calc}

\def\BibTeX{{\rm B\kern-.05em{\sc i\kern-.025em b}\kern-.08em
    T\kern-.1667em\lower.7ex\hbox{E}\kern-.125emX}}

\begin{document}

\title{A Feature Learning Siamese Model for Intelligent Control of the Dynamic Range Compressor}
\thanks{Identify applicable funding agency here. If none, delete this.}

\author{\IEEEauthorblockN{Di Sheng}
\IEEEauthorblockA{\textit{Center for Digital Music} \\
\textit{Queen Mary University of London}\\
London, UK \\
d.sheng@qmul.ac.uk}
\and
\IEEEauthorblockN{Gy{\"o}rgy Fazekas}
\IEEEauthorblockA{\textit{Center for Digital Music} \\
\textit{Queen Mary University of London}\\
London, UK \\
g.fazekas@qmul.ac.uk}
}
\maketitle

\begin{abstract}
% In order to build an intelligent control system for dynamic range compressor (DRC), this paper proposed a siamese structure of DNN model that targets to learn the DRC's characteristics. We have proposed and tested several model designs for the siamese structure and compared the prediction results across them. Since the DNN model is designed to learn a feature embedding, it is also compared with the handcrafted features as we designed in the previous research. The evaluation of the relations between the hyperparameters of DNN and DRC parameters are also provided. This model is able to produce a universal feature embedding that is capable of predicting multiple DRC parameters simultaneously, which is a significant improvement from our previous research. The feature embedding shows a better performance than the handcrafted audio features when we predicting DRC parameters for both mono-instrument audio loops and polyphonic music pieces.
%
In this paper, a siamese DNN model is proposed to learn the characteristics of the audio dynamic range compressor (DRC). This facilitates an intelligent control system that uses audio examples to configure the DRC, a widely used non-linear audio signal conditioning technique in the areas of music production, speech communication and broadcasting. Several alternative siamese DNN architectures are proposed to learn feature embeddings that can characterise subtle effects due to dynamic range compression. These models are compared with each other as well as handcrafted features proposed in previous work. The evaluation of the relations between the hyperparameters of DNN and DRC parameters are also provided. The best model is able to produce a universal feature embedding that is capable of predicting multiple DRC parameters simultaneously, which is a significant improvement from our previous research. The feature embedding shows better performance than handcrafted audio features when predicting DRC parameters for both mono-instrument audio loops and polyphonic music pieces.
\end{abstract}

%\begin{IEEEkeywords}
%Audio Signal Processing, Intelligent Production, DNN, Dynamic Range Compressor, Audio Effects.
%\end{IEEEkeywords}

\section{Introduction}

% GF: This intro is quite similar to all the other papers. This has to be much more focussed on signal processing problems and deep learning for IJCNN. I'll write a few sentences that you can expand later. 

Deep Neural Networks, particularly Convolutional Neural Networks (CNN) have become exceptionally successful in a wide variety of visual object recognition and classification tasks \cite{krizhevsky2012imagenet}. The reasons for this are now well understood. For instance, CNNs can learn filters corresponding to increasingly complex shapes in the target image hence becoming successful at classifying or labelling images. In the domain of audio, the application of CNNs have also proved successful in several tasks, including audio labelling and similarity estimation. The input representation is usually a time-frequency image, e.g. Fourier or Mel-spectrogram \cite{pons2017timbre}\cite{choi2016automatic}\cite{ullrich2014boundary}, but increasingly, raw audio samples are used as well \cite{ardila2016audio}\cite{dieleman2014end}. The reason for the success of these approaches is less straightforward to see, because there is poor analogy between shapes or objects in images, and events, such as notes or chords in audio recordings. Audio events are typically distributed in frequency, e.g., the recording of a note played on an instrument and its harmonic partials activate discontinuous bands along the frequency axis. The problem becomes more acute when audio events overlap. 

Finding an appropriate input representation and designing a neural network suitable for recognising very specific aspects of an audio signal is also a difficult and generally unsolved challenge. Standard approaches work well for common audio classification problems, but if the task becomes focussed on a specific aspect of audio that is often obscured by other large varying signal attributes, we may need to adopt different input representations, network structures and training methods to develop a successful solution. 
For instance, the dynamic range of an audio signal may be characterised by features such as the crest factor \cite{giannoulis2013parameter} and also correlate with note attack and release times. These are measurable in a single note recording but become obscured by overlapping note events and other changes in complex real-world recordings. 

In previous work, we designed an intelligent control system targeting the dynamic range compressor (DRC). This uses an audio example as a reference to estimate DRC parameters that bring a signal closer to the reference \cite{sheng2016automatic}. The system thus takes two inputs, a reference audio and the input signal that needs to be processed. Handcrafted audio features are extracted and fed to a trained machine learning model, which predicts the parameters of DRC and make the output, the compressed input audio, sound as close as possible to the reference. We have designed four sets of handcrafted features that are corresponding to individual DRC parameters: threshold, ratio, attack time and release time  \cite{sheng2018feature:b}. Due to the fact that we used different feature sets for each DRC parameter, we need to train four individual regression models. A generic feature set to predict all parameters would therefore be a great benefit. In addition to the above mentioned drawbacks, most of the handcrafted features are based on note envelope structures. For simple audio material such as isolated notes, it is easy to extract envelopes, however for more complex audio, for instance, audio {\em loops}\footnote{A {\em loop} is a short snippet of musical audio that may be tiled and repeated seamlessly to provide accompaniment.} with overlapping note events, more complex algorithms like NMF and onset event detection are required, to separately analyse note events in a loop \cite{sheng2018feature}. It is still relatively straightforward to extract notes from loops, but polyphonic music brings additional difficulties to the problem and may increase the computational cost dramatically. For instance, overlapping note events may have different duration and timbre, making them difficult to identify. For these reasons, we assume a deep learning model can be beneficial in our task. Deep neural networks can learn complex nonlinear relations between input and output signals. We can design an appropriate audio input representation, and make the model learn a generic feature embedding for all four parameters. Meanwhile, having one trained model to generate features will reduce the computational complexity. The model also has the potential capability to generate efficient features regardless of audio materials. The focus of this paper is to design a feature learning model, so that it will be possible to compare the efficiency between the trained features and handcrafted features. We use the features in conjunction with a conventional regression model, as opposed to end-to-end learning, for sake of reproducibility and easier comparison with previous work. %The model can be designed as a regressor, which is out of the scope of this paper and will be investigated in the following research.   

To achieve our goal, we proposed a siamese structure with the feature embeddings formed by the difference between the output of two branches. We test several architectures within this two branch framework, aiming to learn highly specialised audio features that are invariant to large variations in several other attributes. We first evaluate baseline designs, then we tune them to the task at hand given the observations. The rest of the paper is organised as follows: Section II provides the essential background. Several potentially suitable DNN model designs from the literature are adopted and evaluated in Section III. The models are tuned to our specific problem by altering the model structures and hyperparameters in Section IV. In Section V, we test the most suitable model on larger scale and more complex dataset. Conclusion and future work are provided in Section VI.

%GF: Insert somewhere a sentence about the main contributions of the paper: I think these are 1) the specific use of the siamese structure with the embedding formed by the difference between the output of two branches, and 2) testing several promising architectures for learning highly specialised audio features that are invariant to large variations in several other attributes. 

\section{Background}
\label{background}
Audio effects can be considered signal processing operations applied by audio engineers during music mixing and production. They play an essential role in shaping a desirable sound. Casual users often find it hard to configure the diverse parameters of effects, because they typically have limited understanding of the underlining signal processing. Amateur producers, hobbyists and musicians often describe a desired effect by providing an example, e.g., a name of a style, artist or song as a reference, rather than articulating specific parameters or features, e.g., stating that they require a short attack \cite{mcgrath2016making}. This motivates us to build an intelligent control system using a reference audio. 

Dynamic range compressor is an essential effect in many audio production use cases. DRC operates by applying gain reduction while the input signal energy exceeds a configurable threshold for a set amount of time. It is a highly non-linear effect, which has been the interest of many researchers in intelligent audio production \cite{deman2017ten}. A tutorial on the underlining signal processing of digital DRC is provided in \cite{giannoulis2012digital}. Previous research on intelligent DRC control is using statistical features \cite{hilsamer2014statistical}\cite{ma2015intelligent}. Recent works introduce machine learning and deep learning models \cite{sheng2016automatic}\cite{mimilakis2016deep}.

As discussed earlier, DNNs are increasingly used in audio signal processing and shown outstanding performance in speech recognition \cite{palaz2015convolutional}, music genre classification \cite{li2010automatic} and onset event detection \cite{schluter2013musical} for example. There are attempts of using DNNs for intelligent music production too \cite{mimilakis2016deep,martinez2018end}. The former work uses DNN as a predictor to reproduce a gain factor, while the latter is a generative model that simulates an audio effect in fixed configuration. Since there is no neural network that performs a non-linear signal processing and also have the same parameter control as a conventional DSP algorithm, we use neural networks to learn a feature embedding which is then used for parameter prediction rather than developing an end-to-end model. 

Effective recognition of specific patterns or variations in temporal or spacial data greatly benefits from system characteristics such as shift and translation invariance. This explains the success of CNNs in image and audio processing. There are two primary ways to use CNN for audio related tasks. Since CNN is originally designed for visual data, i.e. 2D signal per channel, we can transform the 1D audio signal into a 2D time-frequency representation and apply 2D CNN directly. Popular representations include the spectrogram \cite{pons2017timbre} and the perceptually motivated Mel-spectrogram \cite{choi2016automatic}\cite{ullrich2014boundary}. Regarding time-frequency representations as equivalent to an image remains a question, even if this has proved to be powerful in several tasks including music classification and tagging. For instance, images can be shifted in the y-axis and the information remains the same, but a shift in the frequency axis will have different implications for audio. Many research shows that given raw audio input, i.e., a uniformly sampled digital signal, it is possible for the model to learn an appropriate hierarchical representation \cite{cakir2018end}. Therefore, to reduce information loss during preprocessing, many recent works apply raw audio as the input for the CNN. Normally, using raw audio involves presenting a relatively long 1D signal. Many researchers use large convolution filter size, e.g. 10-20ms (441-882 samples with a sample rate of 44100Hz) \cite{ardila2016audio}\cite{dieleman2014end}. There are also "sample level" networks which use a small filter size, e.g. 3 samples \cite{lee2017sample} and achieve a comparable level of accuracy in music tagging tasks. Since our problem would require a model to learn multiple parameters of an audio signal processing task, we need the model to be able to capture features at different scales. Therefore the multi-kernel model approach \cite{dieleman2013multiscale}\cite{pons2017end} designed to enhance the versatility of the model would be a benefit in our task. 

Since our problem also requires the model to pay attention to subtle changes in an audio signal, such as the changes in note attack times and to learn features related to them, we consider to use a siamese network \cite{koch2015siamese}. The siamese model is a structure that contains two or more identical subnetworks. It is an appropriate structure when a model needs more than one input or branch, and all inputs are from the same domain. This structure is powerful, especially when the multiple inputs are similar or linked in a certain relation \cite{yu2016sketch}. In our case, the inputs are the audio before and after dynamic range compression. Similar structures may also be used in audio applications for feature learning across different domains \cite{yang2018triplet}.

\section{Model Design}
\label{method}

\begin{figure}[h]
  \centering
  \includegraphics[width=\columnwidth]{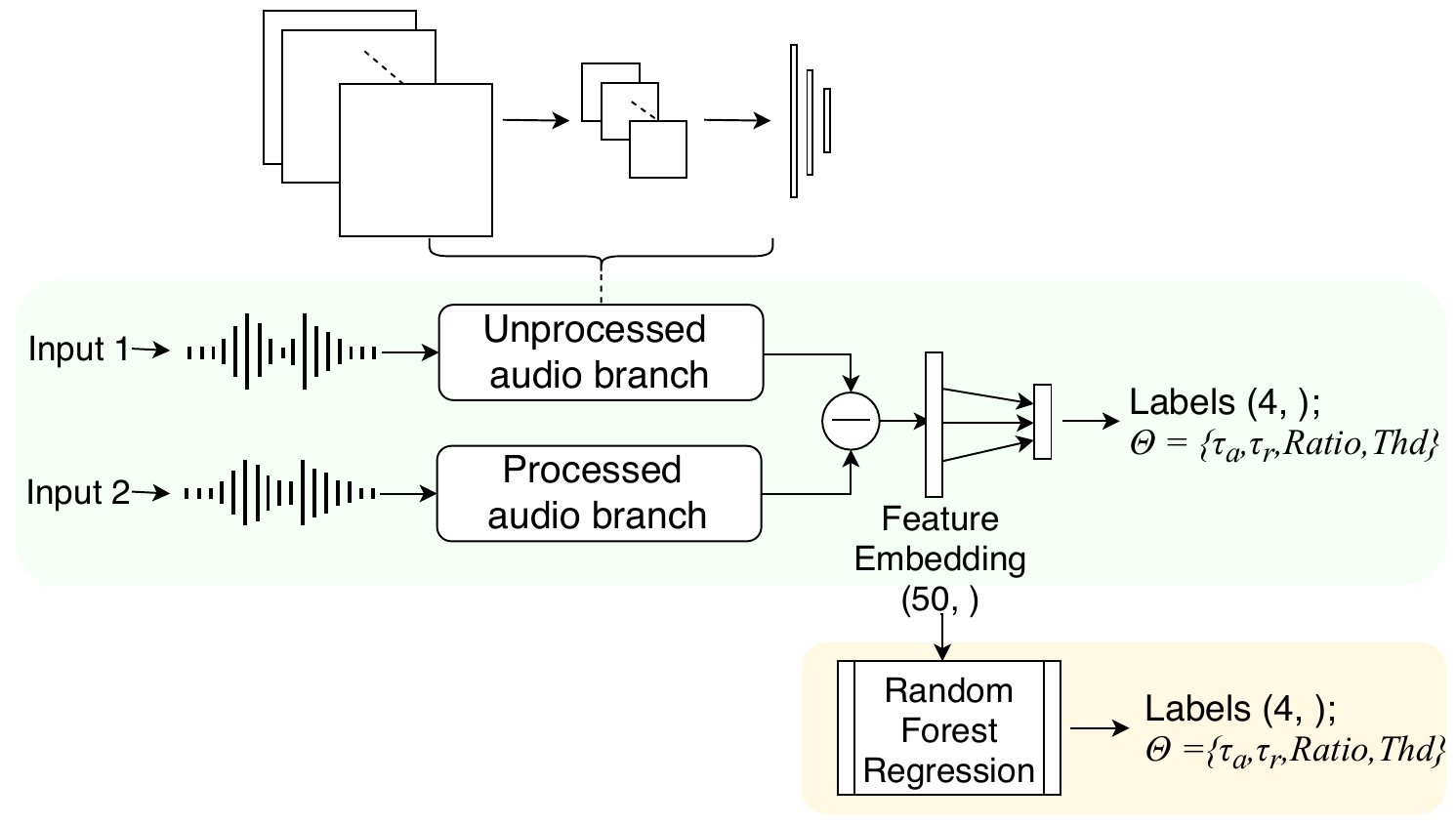}
  \caption{Workflow for the designed system: it contains a twin-siamese DNN model for feature learning with the learning target of DRC parameters, and a random forest regressor trained by DNN feature embedding for parameter prediction. Details of the training process is given in Section \ref{method}.}
  \label{system} %% label for entire figure
\end{figure}

In this research, our purpose is to learn a feature embedding that represents all the characteristics of the DRC. The system uses two audio inputs, an unprocessed input and a compressed audio as reference. During the training process, the latter is a compressed version of the former, therefore, we have access to the ground truth of compression parameters. The dataset with ground truth are divided into training and testing set, therefore, for both training and testing sets, there will be ground truth to evaluate. The evaluation results in this and the following sections are the prediction accuracy of the testing set. In real world scenario, a model performs well if it can still deliver reasonable performance when the reference is no longer generated from the unprocessed input c.f. \cite{sheng2016automatic}. The analysis of this situation is out of the scope of this paper. We focus on learning an optimal feature embedding for a specific audio domain. Figure \ref{system} outlines the training and prediction process. We denote audio as \textit{Input 1} (unprocessed audio), \textit{Input 2} (processed audio) and the ground truth parameters, i.e. training target, as \textit{Labels}. The details of the dataset is provided in Section \ref{data_details}. We have 4 parameters in this research, attack time, release time, ratio, and threshold, denoted $\Theta=\{\tau_a,\tau_r,Ratio,Thd\}$.

\vspace{-8pt}

Two audio files will be used as the inputs for the siamese model shown in Fig. \ref{system}, where they are described as \textit{Unprocessed branch} and \textit{Processed audio branch}. These identical branches contain convolutional layers as well as dense layers. Their outputs are considered feature embeddings. We will introduce three model designs in the following paragraph. To learn the characteristics of an audio signal, we can preprocess it to obtain a time-frequency representation, or use raw audio samples as input. We have two model designs corresponding to these two types of inputs signals. Besides, we also consider the multi-kernel approach \cite{pons2017end} to enable learning features at multiple time and frequency scales. We merge the output of the two branches using subtraction, followed by a fully connected dense layer. Mean squared error is used as loss function for training as the training targets are the parameters $\Theta$ of the DRC. Finally, we update the learning rate adaptively using Adadelta \cite{zeiler2012adadelta}. The trained feature embeddings are then used as features to train a random forest regression model. It follows the same procedure of our previous work \cite{sheng2016automatic}. We did not design the model as a predictor directly, but it is possible to plug a DNN regressor after the feature embedding layer. At the current stage of the research, we prefer to focus on enabling the model to learn DRC related features well. We would also like to compare the results with our previous handcrafted features. Using the random forest model will therefore provide a more trustworthy comparison.

\subsection{Model designs for the siamese branches}
\noindent\textit{Model 1. CNN structure}

%GF: What problem does this model solves? Why is it proposed?

The first model design for both identical branches is the classical CNN structure \cite{krizhevsky2012imagenet} widely used in image and audio signal analysis. CNNs provide the state-of-the-art in relevant tasks including onset event detection \cite{schluter2014improved} and music boundary learning \cite{ullrich2014boundary}. The audio segments that are affected at the DRC onset, i.e., when the signal exceeds the configurable threshold, can be considered a boundary which represents a change point within the audio. Multiple aspects of the audio are altered depending on the parameter settings. We assume the CNN is able to learn the DRC characteristics. We use Mel-spectrogram input, a good reduced representation of timbre information. We use a seven layer CNN, which consists of five convolutional layers, five max-pooling layers and a dropout rate of 0.1. It is followed by two dense layers. The model summary is provided in Table. \ref{model1}.

\begin{table}[h]
\centering
\begin{tabular}{|c|}
\hline
Input: Mel-spectrogram (128,690,1)                                                  \\ \hline
\hline
\begin{tabular}[c]{@{}c@{}}Conv2D: 3*3*10\\ MaxPool2D: 2*2\\ DropOut: 0.1\end{tabular} \\ \hline
\begin{tabular}[c]{@{}c@{}}Conv2D: 3*3*15\\ MaxPool2D: 2*2\\ DropOut: 0.1\end{tabular} \\ \hline
\begin{tabular}[c]{@{}c@{}}Conv2D: 3*3*15\\ MaxPool2D: 2*2\\ DropOut: 0.1\end{tabular} \\ \hline
\begin{tabular}[c]{@{}c@{}}Conv2D: 3*3*20\\ MaxPool2D: 2*2\\ DropOut: 0.1\end{tabular} \\ \hline
\begin{tabular}[c]{@{}c@{}}Conv2D: 3*3*20\\ MaxPool2D: 2*2\\ DropOut: 0.1\end{tabular} \\ \hline
\begin{tabular}[c]{@{}c@{}}Flatten\\ Dense(feature embedding layer): 50\\ Dense: num\_para \end{tabular}                         \\ \hline
\hline
Output: Parameters                                                                   \\ \hline
\end{tabular}
\vspace{2pt}
\caption{Model summary for the CNN structure, i.e. Model 1.}
\label{model1}
\end{table}

\vspace{-5pt}
\noindent\textit{Model 2. Sample level CNN for waveform input}

%GF: What problem does this model solves? How is it different from the previous one? Why is it proposed?
The second model structure uses time domain audio samples as input. As we mentioned before, the common time-frequency representation cannot be considered exactly equivalent to an image, therefore we propose using audio samples directly as input. Since some aspects of DRC operation may be better characterised in the time domain, i.e., at the audio sample level, we used sample-level small filters and follow the model design in \cite{lee2017sample}, where the proposed network contains seven 1D convolutional layers, batch-normalisation, and six layers of max-pooling. This front-end is then followed by two residual layers and two dense layers. We use residual layers to avoid the vanishing gradient problem without introducing too many layers \cite{he2016deep}. A summary of this model is provided in Table \ref{model2}. Some of the convolutional layers are duplicated, we use ``*2'' to represent two groups of layers with the same settings. ``L1''-``L4'' are the notations for back end layers.

% The models taking raw audio as input have provided equivalent performance with the ones using time-frequency representations as input, c.f. Section \ref{background}.

%\Todo{explain why}

\begin{table}[ht]
\centering
\begin{tabular}{|c|c|c|}
\hline
\multicolumn{3}{|c|}{Input: Waveform (44100,1)}                                                              \\ 
\midrule
\hline
\multirow{9}{*}{\begin{tabular}[c]{@{}c@{}} \\ Front\\ End\\ Network\end{tabular}} & \multicolumn{2}{c|}{\begin{tabular}[c]{@{}c@{}}Conv1D: 3*1*64\\ Batch Normalisation\end{tabular}}           \\ \cline{2-3}
&\begin{tabular}[c]{@{}c@{}}Conv1D: 3*1*64\\ Batch Normalisation\\ MaxPool1D: 3*1\end{tabular}      & *2      \\ \cline{2-3}
&\begin{tabular}[c]{@{}c@{}}Conv1D: 3*1*128\\ Batch Normalisation\\ MaxPool1D: 3*1\end{tabular}      & *2      \\ \cline{2-3}
&\begin{tabular}[c]{@{}c@{}}Conv1D: 3*1*256\\ Batch Normalisation\\ MaxPool1D: 3*1\end{tabular}      & *2      \\ \cline{2-3}
&\multicolumn{2}{c|}{Flatten, Dimension\_expand}                                                             \\ \hline
\multirow{9}{*}{\begin{tabular}[c]{@{}c@{}} \\ Back\\ End\\ Network\end{tabular}}&\begin{tabular}[c]{@{}c@{}}Conv2D: 7*256*512\\ Batch Normalisation\end{tabular}                    & L1;     \\ \cline{2-3}
&\begin{tabular}[c]{@{}c@{}}Conv2D: 7*256*512\\ Batch Normalisation\end{tabular}                    & L2;     \\ \cline{2-3}
& Add (L1, L2)                                                                                       & L3;     \\ \cline{2-3}
&\begin{tabular}[c]{@{}c@{}}Conv2D: 7*256*512\\ Batch Normalisation\end{tabular}                    & L4;     \\ \cline{2-3}
&\multicolumn{2}{c|}{Add (L3, L4)}                                                                           \\ \cline{2-3}
&\multicolumn{2}{c|}{\begin{tabular}[c]{@{}c@{}}Global Pooling;\\ Dense(feature embedding layer): 50\\ Dense: num\_para\end{tabular}} \\ \hline
\midrule

\multicolumn{3}{|c|}{Output: Parameters}                                                                      \\ \hline

\end{tabular}
\vspace{2pt}
\caption{Model summery for the waveform structure, i.e. Model 2. It can be separated to front-end and back-end network, where the front-end is a combination of sample level 1D Conv layers and the back-end consists of two residual layers.}
\label{model2}
\end{table}

\vspace{-8pt}
\noindent\textit{Model 3. Multi-kernel CNN structure}

\begin{figure*}[ht]
  \centering
  \includegraphics[width=1.8\columnwidth]{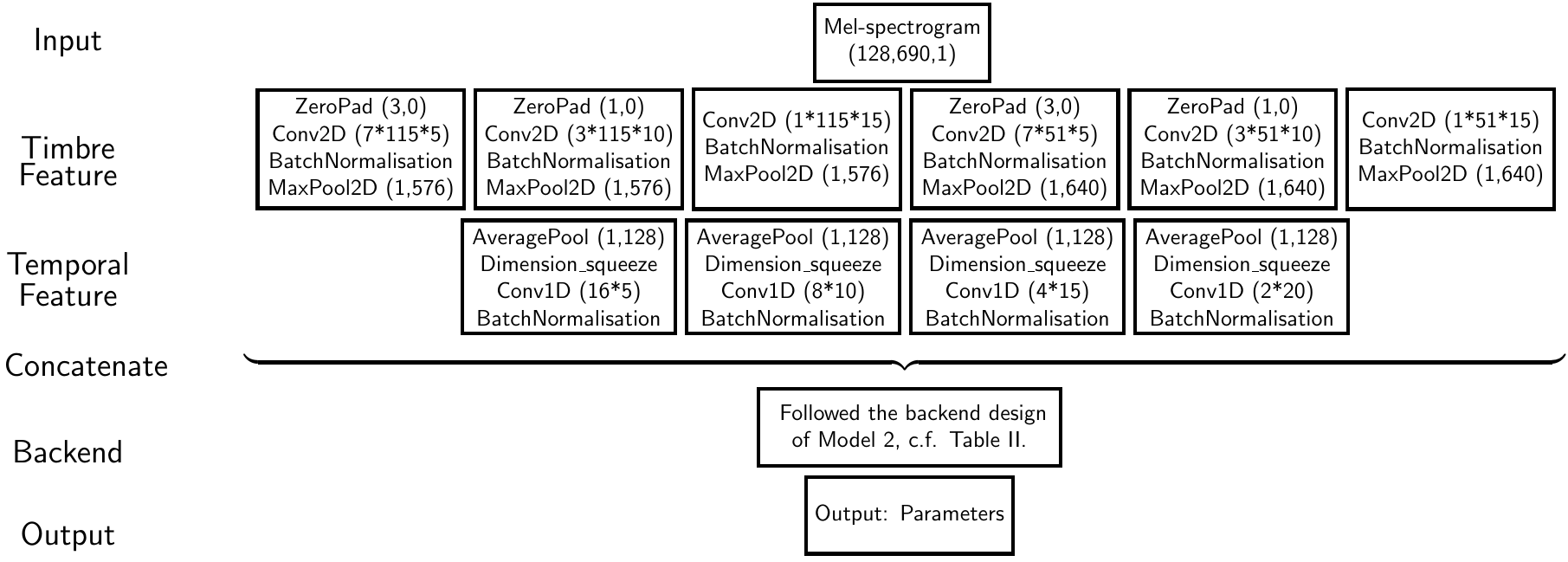}
  \caption{Model summery for the multi-kernel structure, i.e. Model 3. The front end network concatenates 11 Conv layers with different kernel shapes. The back end network is two layers of residual layers, which is the same as Table \ref{model2}. }
  \label{model3} %% label for entire figure
\end{figure*}

%GF: What problem does this model solves? How is it different from the previous one? Why is it proposed?
Our task is to make this siamese model learn multiple aspects of DRC, i.e., changes that happen at different time scales, as well as at different magnitudes, for example, due to changes in attack time vs. ratio and threshold. We consider to use the multi-kernel model construction proposed in \cite{pons2017end}. This model is designed to capture the audio features at multiple scales at the same time, which fit our purpose to observe audio characteristics over different decision horizons. It could be especially useful for our problem because we want our model to learn four aspects of DRC at the same. The model with only one kernel size might neglect certain features of the audio. The model applies several temporal kernels as well as frequency domain (timbre) kernels as illustrated in Fig. \ref{model3}. Similarly to the previous model, this structure uses Mel-spectrogram input too. It applies six different kernel shapes with 2D convolutional layers, four different kernel shapes with 1D convolutional layers and concatenate the outputs of all ten layers together as input to the back-end network, which has two residual convolutional layers and two dense layers as in Model 2.

\subsection{Dataset description}
\label{data_details}
In our experiments, we generate 64 guitar loops and drum loops using the Apple Loops\footnote[1]{\url{https://support.apple.com/kb/PH13426}} library. We compress them using different DRC parameter settings to produce the following datasets:

For $DS_1$ to $DS_4$, the audio files are compressed with only one parameter changing, while the others are fixed. For $DM_1$ and $DM_2$, each audio file is compressed while two parameters are changing at the same time. Take  $DM_1$ as an example, each audio file will be compressed by 8 threshold settings and 8 ratio settings. It will produce 64 compressed audio for each raw audio loop. For instance, $guitar_1$ is compressed using $Thd$ [10.0dB, 14.8dB, 19.6dB, 24.4dB, 29.2dB, 34.0dB, 38.8dB, 43.6dB], $guitar_2$ is compressed using $Thd$ [10.6dB, 15.4dB, 20.2dB, 25.0dB, 29.8dB, 34.6dB, 39.4dB, 44.0dB], and so on. For each audio file, the compression threshold grid is 4.8dB, but the combined set yields a finer grid, 0.6dB as noted in the table. The default value for each parameter is $Thd=-37.5dB$, $Ratio=2:1$, $\theta_a=5ms$, $\theta_r=200ms$.

\begin{table}[ht]
\centering
\begin{tabular}{l|l|l}
\toprule
                   & dataset generation                                                                                                               & dataset size               \\ \midrule
$DS_1$ & $Thd$: 0 to 49dB with step of1dB                                                                                             & \begin{tabular}[c]{@{}l@{}}guitar: 65*50;\\ drum: 64*50\end{tabular} \\ \hline
$DS_2$ & $Ratio$: 0 to 20 with step of 0.4                                                                                                  & \begin{tabular}[c]{@{}l@{}}guitar: 65*50;\\ drum: 64*50\end{tabular} \\ \hline
$DS_3$ & $\tau_a$: 1 to 100ms with step of 2ms                                                                                         & \begin{tabular}[c]{@{}l@{}}guitar: 65*50;\\ drum: 64*50\end{tabular} \\ \hline
$DS_4$ & $\tau_r$: 10 to 1000ms with step of 20ms                                                                                     & \begin{tabular}[c]{@{}l@{}}guitar: 65*50;\\ drum: 64*50\end{tabular} \\ \midrule
$DM_1$  & \begin{tabular}[c]{@{}l@{}}$Thd$: 10 to 47.8dB with step of 0.6dB \\ $Ratio$: 1 to 19.9 with step of 0.3\end{tabular}          & \begin{tabular}[c]{@{}l@{}}guitar: 65*64;\\ drum: 64*64\end{tabular} \\ \hline
$DM_2$  & \begin{tabular}[c]{@{}l@{}}$\tau_a$: 1 to 95.5ms with step of 1.5ms\\ $\tau_r$:10 to 955ms with step of 15ms\end{tabular} & \begin{tabular}[c]{@{}l@{}}guitar: 65*64;\\ drum: 64*64\end{tabular} \\ \bottomrule
\end{tabular}
\vspace{2pt}
\caption{Dataset details for two instruments}
\label{dataset_table}
\end{table}

In the following subsection, models for one and two parameters are trained and tested based on these datasets. A more complex dataset is used in Section \ref{evaluation}.

\subsection{Evaluation of different model designs}

All models are trained using a small batch size of 8, while 15\% of the data are used as validation set. We monitored the validation error to avoid overfitting to the training data. After the model is trained, the feature embedding is generated using the trained networks. The feature embeddings are then used to train a random forest regression model, where we split the data into 80\% training and 20\% testing. We split the audio dataset randomly 50 times. The averages of the test prediction mean absolute errors (MAE) are reported in Table \ref{initial}.

The DNN models are able to produce similar results for $Thd$ and $Ratio$. However, the prediction error of $\tau_a$ and $\tau_r$ are relatively high compared to handcrafted features. Since we did not preprocess input audio to emphasise any of the DRC's effect, the model would react better when a specific DRC parameter has a more significant influence on the raw audio, i.e. $Thd$ and $Ratio$. Moreover, handcrafted features are tuned specifically to extract information from a temporal region of the audio where a certain parameter is the most effective. Comparing the two types of models, using raw audio as input provides better performance in 6 out of 8 cases for single parameter prediction. Many factors can lead to this result. First of all, the frame size we used for the Mel-spectrogram is 512, which is relatively large given that our problem focuses on small transient times. Secondly, Mel-spectrogram works well when the aim is to retrieving high-level musical information, but it may smear useful spectral information that are important in our problem. Given these observations, some improvements on the network structures and hyperparameters are described in the next section.

\begin{table}[h]
\centering
\begin{tabular}{c|c|c|c|c|c}
\toprule
\multicolumn{2}{c|}{}             & $Thd$ & $Ratio$ & $\tau_a$  & $\tau_r$  \\ \midrule
\multirow{4}{*}{Guitar} & Model 1  & 1.781dB   & 0.657 & 4.338ms & 35.589ms \\ \cline{2-6} 
                        & Model 2  & 1.206dB   & 0.751 & 3.192ms & 32.893ms \\ \cline{2-6} 
                        & Model 3  & 1.034dB  & 1.009 & 3.273ms & 71.288ms \\ \cline{2-6} 
                        & Handcrafted & \textbf{0.903dB}   & \textbf{0.623} & \textbf{0.845ms} & \textbf{10.442ms} \\ \midrule
\multirow{4}{*}{Drum}   & Model 1  & 2.994dB   & 0.961 & 3.829ms & 58.394ms \\ \cline{2-6} 
                        & Model 2  & 2.627dB   & 0.932 & 3.480ms & 43.668ms \\ \cline{2-6} 
                        & Model 3  & 2.953dB   & 1.218 & 7.694ms & 93.064ms \\ \cline{2-6} 
                        & Handcrafted & \textbf{0.915dB}   & \textbf{0.655} & \textbf{1.194ms} & \textbf{12.714ms} \\ \bottomrule
\end{tabular}
\vspace{2pt}
\caption{Prediction MAE for regression model using feature embeddings learnt from each DNN as well as handcrafted features, when predicting individual parameters of the DRC.}
\label{initial}
\end{table}

The results are not yet encouraging when using the trained features to predict individual DRC parameters, in contrast, improvements are shown when we train the model to learn two parameters simultaneously. The training data for the two parameters model are $DM_1$ and $DM_2$, which is larger than the single parameter training set. The results are shown in Table \ref{1st}. Comparing with Table \ref{initial}, the model produces better results, especially for release time. For $Thd$, $Ratio$, and $\tau_r$, the DNN model is able to yield a better performance than handcrafted features. Model 2 is the best performing model in this experiment. This model provides the best performance for 4 out of 8 cases. For reference, the range of each parameters are 49dB for $Thd$, 19 for $Ratio$, 99ms for $\tau_a$ and 999ms for $\tau_r$. The larger range of release time results in a higher prediction error compared with the other parameters.

\begin{table}[h]
\centering
\begin{tabular}{c|c|c|c|c|c}
\toprule
\multicolumn{2}{c|}{}             & \multicolumn{2}{c|}{$Thd$, $Ratio$} & \multicolumn{2}{c}{$\tau_a$, $\tau_r$}   \\  \midrule
\multirow{4}{*}{Guitar} & Model 1  & \multicolumn{2}{c|}{1.854dB, 0.529}                 & \multicolumn{2}{c}{2.081ms, 20.184ms}                  \\ \cline{2-6} 
                        & Model 2  & \multicolumn{2}{c|}{1.666dB, \textbf{0.460}}                 & \multicolumn{2}{c}{1.725ms, 18.357ms}                  \\ \cline{2-6} 
                        & Model 3  & \multicolumn{2}{c|}{1.567dB, 0.618}                 & \multicolumn{2}{c}{\textbf{1.565ms}, \textbf{15.588ms} }                \\ \cline{2-6} 

                        & Handcrafted & \multicolumn{2}{c|}{\textbf{0.912dB}, 0.883}                 & \multicolumn{2}{c}{2.100ms, 25.079ms}                  \\ \midrule
\multirow{4}{*}{Drum}   & Model 1  & \multicolumn{2}{c|}{1.810dB, 0.800}   & \multicolumn{2}{c}{5.061ms, 27.005ms} \\ \cline{2-6} 
                        & Model 2  & \multicolumn{2}{c|}{\textbf{1.112dB, 0.391}}   & \multicolumn{2}{c}{4.506ms, \textbf{13.609ms}} \\ \cline{2-6} 
                        & Model 3  & \multicolumn{2}{c|}{2.170dB, 0.782}                 & \multicolumn{2}{c}{6.463ms, 21.427ms}  \\ \cline{2-6} 
                        & Handcrafted & \multicolumn{2}{c|}{3.233dB, 0.684}   & \multicolumn{2}{c}{\textbf{2.354ms}, 14.980ms} \\ \bottomrule
\end{tabular}
\vspace{2pt}
\caption{Prediction MAE for regression model using feature embeddings learnt from each DNN as well as handcrafted features, when the model is jointly predict two DRC parameters.}
\label{1st}
\end{table}

%GF: Before introducing each model in this section I would first explain the specific recognition task (i.e. say why DRC parameters are difficult to estimate, and/or discuss the specificity of handcrafted features) then form a hypothesis of why a particular model might be good to solve the outlined problem. Between each model I would indicate the main conceptual difference and possible deficiency of the previous model. I'd adopt a uniform structure for this line of argument.

\vspace{-10pt}
\section{Model Tuning}

In the previous section, we explored several model designs from the literature. In this section, we improve the networks to better fit our task, along with the analysis of the model components' impact on the results. The following subsections discuss each model respectively. 

\subsection{Improvements on Model 1}
\label{model1_ipv}

First, we aim to improve the performance of Model 1. There are several reasons to use time-frequency representation as input: the data size is reduced; the ability to control the frequency and time resolution; using 2D convolution will give us more choices of kernel size; 

The previous settings for the hyperparameters are used directly as reported in \cite{pons2017end}\cite{lee2017sample}. Next, we optimise the network structure and hyper parameters. As discussed in Section \ref{background}, many factors will influence the performance. Here, we considered three factors: whether using Mel-spectrogram or spectrogram, the kernel shape of the model and finally the frame length of the STFT. We can assume that the Mel-spectrogram smeares useful information and this leads to a poor performance. Therefore, using spectrogram alone can be assumed to improve the performance. Due to the fact that our problem requires focussing on a short transient time at some point, as well as the sample-level Model 2 shows a better performance in the previous experiment, we assume that a short time frame, i.e., a better time resolution and a smaller kernel size will also improve the performance. Based on these two assumptions, we conduct the following experiments.

The first exploratory analyses are designed to improve the model, therefore we do not conduct a thorough experiments for all the datasets. We use the drum dataset: $DS_3$, $DS_4$ and $DM_2$ in this Section, because it shows in Table \ref{initial} that prediction of attack/release time is more difficult for DNN models (large performance gap). The predicted mean absolute errors (MAE) are provided in Table \ref{test1}. The first experiment aims to select the most suitable input signal format. We use the same frequency resolution for both representations. The prediction error shows a large improvement. Especially for release time, it reaches a similar performance to Model 2. All cases exceed the performance of Model 1 too. We can conclude that spectrogram is a more suitable representation for this task and this model.

\begin{table}[h]
\centering
\begin{tabular}{c|c|c}
\toprule
\diagbox{Para(ms)}{Input signal} & Melgram & Spectrogram \\ \midrule 
$\tau_a$                       & 3.829   & \textbf{2.415}       \\ \midrule
$\tau_r$                     & 58.394  & \textbf{33.085}      \\ \midrule
\multirow{2}{*}{Joint prediction ($\tau_a$/$\tau_r$)} & 5.061   & \textbf{4.656}       \\ 
                                  & 27.005  & \textbf{17.781}     \\ \bottomrule
\end{tabular}
\vspace{2pt}
\caption{Prediction MAE when changing input representations}
\label{test1}
\end{table}

In the second experiment, we investigate the time frame length of spectrograms. The results in Table \ref{time_frame} provides a clear trend that with the decrease of frame size, the prediction error drops. We did not progress the experiment with frame size smaller than 128 samples because this would yield unreasonably low frequency resolution.

\begin{table}[h]
\centering
\begin{tabular}{c|c|c|c}
\toprule
\diagbox{Para (ms)}{Time frame length}  & 512    & 256    & 128    \\ \midrule
$\tau_a$                       & 2.415  & 2.197  & \textbf{2.042}  \\ \midrule
$\tau_r$                     & 33.085 & 30.546 & \textbf{26.698} \\ \midrule
\multirow{2}{*}{Joint prediction ($\tau_a$/$\tau_r$)} & 4.656  & 4.271  & \textbf{3.141}  \\  
                                  & 17.781 & 16.846 & \textbf{15.752} \\ \bottomrule
\end{tabular}
\vspace{2pt}
\caption{Prediction MAE when changing frame size for spectrogram}
\label{time_frame}
\end{table}

The third experiment is conducted while we alter the kernel size of the model. The original design is using five 2D convolutional layers with a 3 by 3 kernel. Since we need to capture audio features in multiple feature dimensions, we will try to alter kernel sizes and combinations. We reduce the depth of the 2D layers and increase the depth of 1D convolutional layers at the same time. Except for the release time results, the other performances did not show significant improvement. For simplicity, we keep the five convolutional layers with 3*3 kernels in further experiments.

\begin{table}[h]
\centering
\begin{tabular}{c|c|c|c}
\toprule
\diagbox{Para (ms)}{Kernel size}   & 5(3*3)     & 4(3*3)+1(1*3)    & 3(3*3)+2(1*3)    \\ \midrule
$\tau_a$                       & 2.042  & 1.966  & \textbf{1.962}  \\ \midrule
$\tau_r$                      & 26.698 & 21.173 & \textbf{18.491} \\ \midrule
\multirow{2}{*}{Joint prediction ($\tau_a$/$\tau_r$)} & \textbf{3.141}  & 4.232  & 3.941  \\ 
                                  & 15.752 & \textbf{14.436} & 16.791 \\ \bottomrule
\end{tabular}
\vspace{2pt}
\caption{Prediction MAE when changing kernel shapes for Model 1, with different combinations of 2D and 1D Conv layers}
\end{table}

\subsection{Improvement on Model 2}

% Followed by the previous section, tuning the structure and hyperparameters of Model 1 leads to a performance improvement. In many cases, it exceeds the performance of using raw audio signal as input. 
In this section, we explore the improvement for Model 2 using the same datasets. One conclusion we can draw from the previous experiment is that for our problem, having large filter size in the time axis will result in poor performance in predicting {$\tau_a, \tau_r$}. We have also tested the performance on {$Ratio, Thd$} but the improvement is not as significant as the temporal parameters, therefore we did not present them. Apart from the size of the filters, we also alter the number of filters and layers of the network. The prediction errors are provided in Table \ref{model2imp}. However, the results do not improve as significantly as they do for Model 1. Even though the advantage of the sample level CNN is having a fine resolution in time domain, we can achieve and exceed the prediction result by tuning the 2D CNN model and its input time-frequency representation.

\begin{table}[ht]
\centering
\begin{tabular}{c|c|c|c|c}
\toprule

                                 &    Initial     &  $\uparrow$ Filter size                                    &  $\downarrow$ Filter number                                    & $\downarrow$ Layers                                            \\ 
                                 \midrule
$\tau_a$ & 3.480ms & 7.772ms & \textbf{3.227ms} &  3.784ms \\ \midrule
$\tau_r$ & 43.668ms & 48.834ms & \textbf{39.740ms}&  45.030ms \\                                 
                                 \midrule
\begin{tabular}[c]{@{}c@{}}Joint\\prediction\end{tabular}  &
\begin{tabular}[c]{@{}c@{}} 4.506ms\\ \textbf{13.609ms}\end{tabular} & \begin{tabular}[c]{@{}c@{}} 6.683ms\\ 20.486ms\end{tabular} & \begin{tabular}[c]{@{}c@{}} 5.847ms\\ 17.622ms\end{tabular} & \begin{tabular}[c]{@{}c@{}} \textbf{3.864ms}\\ 16.619ms\end{tabular} \\ \bottomrule
\end{tabular}
\vspace{2pt}
\caption{Tuning results for Model 2, when we increase filter size, reduce filter numbers, and reduce layers respectively}
\label{model2imp}
\end{table}

\subsection{Improvement on Model 3}

As it is mentioned in Section \ref{method}, we also consider the multi-kernel structure that proved to be efficient in the music tagging problem \cite{pons2017end}. Since the DRC may impact audio events in the short and long-term and impact different frequencies in a non-linear fashion, it makes sense to use multiple kernels at the same time. We use all six datasets to train the best performing model (Model 1 tuned), Model 3 with default settings, and the Model 3 with tuned hyperparameters. We used all six datasets to provide an overall evaluation for the tuned models. The prediction results are provided in Table \ref{overall_performance}.  

\begin{table}[h]
\centering
\begin{tabular}{c|c|c|c}
\toprule
                             & Model 1 tuned  & Model 3  & Model 3 tuned \\ \midrule
$Thd$                    & 1.543dB  & 2.953dB  & 2.602dB       \\ \midrule
$Ratio$                        & 0.746    & 1.218    & 1.184       \\ \midrule
$\tau_a$                  & 2.024ms  & 7.694ms  & 7.691ms             \\ \midrule
$\tau_r$                 & 26.698ms & 93.064ms & 41.678ms         \\ \midrule
\multirow{4}{*}{Joint prediction} & 1.019dB  & 2.170dB  &  1.351dB      \\  
                             & 0.417    & 0.782    &  0.394
\\ \cline{2-4}
                             & 3.141ms  & 6.463ms  & 3.644ms        \\  
                             & 15.752ms & 21.427ms & 17.688ms         \\ \bottomrule
\end{tabular}
\vspace{2pt}
\caption{Improvement for Model 3, using spectrogram and a reduction of frame size}
\label{overall_performance}
\end{table}

The results for Model 3 are not as good as the best performance reported in Section \ref{model1_ipv}. Based on the conclusion from the previous section, we improve the model by using spectrogram with smaller time frame as input, as well as decreasing the kernel size for the temporal features. These changes do not improve the prediction error rate significantly however. Model 3 combines multiple feature representation layers, therefore there are more trainable parameters compared to Model 1 and Model 2. This leads to growth in complexity and make the training more difficult. It might also be due to the shallowness of the network. The model concatenated multiple layers, but they are all single layers, therefore the depth of the model is 3. This is very shallow compared to the other designs. We will consider to deepen this model in the future work.

\section{Evaluation on simultaneous parameter estimation and polyphonic music data}
\label{evaluation}
In this experiment, we expand the dataset from two parameters changing simultaneously to all four parameters changing together. The data is generated using the same method described earlier and reported in Table \ref{dataset_table}. Changing four parameters together results in a substantial growth of the amount of data, therefore, instead of 8 settings for each parameters, we use 5, i.e., $drum_1$ is compressed using $Thd$: [10.0dB, 18dB, 26dB, 34dB, 42dB], $Ratio$: [1.28:1, 5.12:1, 8.96:1, 12.80:1, 16.64:1], $\tau_a$: [1ms, 21ms, 41ms, 61ms, 81ms], and $\tau_r$: [10ms, 210ms, 410ms, 610ms, 810ms]; $drum_2$ is compressed using $Thd$ [11.0dB, 19dB, 27dB, 35dB, 43dB], $Ratio$: [1.76:1, 5.60:1, 9.44:1, 13.28:1, 17.12:1], $\tau_a$: [3.5ms, 23.5ms, 43.5ms, 63.5ms, 83.5ms], and $\tau_r$: [35ms, 235ms, 435ms, 635ms, 835ms], and so on. The new dataset of size $64*625=40000$ is outlined in Table \ref{new_data}.

\begin{table}[h]
\centering
\begin{tabular}{c|c|c}
\toprule
                   & dataset generation                                                                                                               & dataset size               \\ \midrule
D4P  & \begin{tabular}[c]{@{}c@{}}$Thd$: 10 to 49dB with step of 1dB\\ $Ratio$: 1.28 to 20 with step of 0.48\\ $\tau_a$: 1 to 98.5ms with step of 2.5ms\\ $\tau_r$:10 to 985ms with step of 25ms\end{tabular} &  drum: 64*625 \\ \bottomrule
\end{tabular}
\vspace{2pt}
\caption{Dataset details for changing four parameters together}
\label{new_data}
\end{table}

We compare the prediction results between the regressor trained using handcrafted features and feature embedding learnt by Model 1. The results are given in Table \ref{real}. The predictions of the four parameter model are not as good as the two parameter ones. However, the four parameter model shows its advantage when compared to handcrafted features. When several attributes of the audio are changing at a time, handcrafted features that are designed to measure specific attributes like attack time differences, tend to lose their benefit compared to a neural network. The MAEs for attack and release time have grown, but for reference, the range of the two parameters are 99ms and 999ms respectively. We also outline the percentage of the predicted error over the parameter range in Table \ref{real}.

\begin{table}[h]
\centering
\begin{tabular}{c|c|c}
\toprule
                             & Handcrafted features  & Feature embeddings   \\ \midrule
$Thd$                    & 2.937dB / 7.34\%  & 2.369dB / 5.92\% \\ \midrule
$Ratio$                        & 3.447 / 17.24\%    & 3.265  / 16.33\%   \\ \midrule
$\tau_a$                  & 13.926ms / 14.07\% & 10.868ms  / 10.98\% \\ \midrule
$\tau_r$                 & 120.577ms / 12.07\% & 79.045ms / 7.91\% \\ 
\bottomrule
\end{tabular}
\vspace{2pt}
\caption{Prediction MAE using large scale data and comparing handcrafted features and feature embeddings for predicting four DRC parameters. The percentage of the prediction error over parameter range is also given.}
\label{real}
\end{table}

We also test the model using more complex audio materials, that is, polyphonic music. We randomly select 50 audio segments from the mixed music of the MedleyDB dataset\cite{Bittner14medleydb:a}. We generate 50*625 compressed audio using the method described in Table \ref{new_data}. We use the feature learning model trained on drum loops' features to predict the compression parameters of the mixed audio. The two types of features are handcrafted features and the feature embeddings from the best performing DNN model. The predicted MAEs are reported in Table \ref{mix}. The results from the model trained by feature embeddings are clearly better. This result is reasonable because considering that the programme material is different from what model is tuned to, however, handcrafted features are more depended on it. It is surprising that the prediction MAE for mix audio is better than testing drum loops alone (compare the last column of Table \ref{real} and Table \ref{mix}). We may assume that richer audio content provides a benefit for the DNN model. The results also indicate improved robustness and generalisation of our method. Our future work aims to confirm or refute these observations.

\begin{table}[h]
\centering
\begin{tabular}{c|c|c}
\toprule
                             & Handcrafted features  & Feature embeddings   \\ \midrule
$Thd$                    & 11.585dB  & 1.697dB   \\ \midrule
$Ratio$                        & 5.104     & 2.194     \\ \midrule
$\tau_a$                  & 26.628ms  & 9.873ms   \\ \midrule
$\tau_r$                & 268.420ms & 160.629ms  \\ 
\bottomrule
\end{tabular}
\vspace{2pt}
\caption{Prediction MAE for mixed audio whose feature embeddings are generated using DNN model trained by drum loops.}
\label{mix}
\end{table}

\section{Conclusion}

In this paper, we explored several model designs for an intelligent DRC control system. Handcrafted features failed to provide good prediction when we aim to predict several parameters together. DNN models start to show their advantage compared to handcraft features when we predict two or more parameters. The improvement becomes substantial when predicting all four parameters together. Across the three model designs, the CNN model using a high temporal resolution spectrogram as input shows the best performance. 

In this research, we discovered that the performance would improve significantly when an appropriate time-frequency representation is used as input. Surprisingly, the multi-kernel model does not provide the best performance in this task. This might be due to the complexity of the model, i.e., a large network with more parameters possibly requiring more training data. However, it is still worth considering to use the multi-resolution time-frequency representations as input. When we predicting more parameters together, the performance for two parameters are much better than predicting only one. However, the performance for four parameters drop substantially compared with predicting only two parameters. A possible explanation could relate to the larger grid size when training data for the four parameters. The results may improve using a finer grid and increased training data size. 

In conclusion, the DNN model provides the ability to train one generic model for all parameters of an audio effect. It helps to reduce the limitations of handcrafted features. Further research will be conducted to improve the performance and test our models in a more realistic real-world scenarios, e.g., when applying the system in a professional music recording studio.

%\begin{thebibliography}{00}
%\bibitem{b1} G. Eason, B. Noble, and I. N. Sneddon, ``On certain integrals of Lipschitz-Hankel type involving products of Bessel functions,'' Phil. Trans. Roy. Soc. London, vol. A247, pp. 529--551, April 1955.
%\end{thebibliography}
\bibliographystyle{IEEEtran}
\bibliography{ref}

\end{document}